# Fast versus conventional HAADF-STEM tomography: advantages and challenges

*Hans Vanrompay[1,2], Alexander Skorikov[1,2], Eva Bladt[1,2], Armand Béché[1,2], Bert Freitag[3], Jo Verbeeck[1,2], Sara Bals[1,2†]*

[1]Electron Microscopy for Materials Science (EMAT), University of Antwerp, Groenenborgerlaan 171, 2020 Antwerp, Belgium

[2]NANOlab Center of Excellence, University of Antwerp, Belgium

[3]Thermo Fisher Scientific, Achtseweg Noord 5, 5651 GG Eindhoven, The Netherlands

[†]Correspondence to: sara.bals@uantwerpen.be

Abstract

Electron tomography is a widely used experimental technique for analyzing nanometer-scale structures of a large variety of materials in three dimensions. Unfortunately, the acquisition of conventional electron tomography tilt series can easily take up one hour or more, depending on the complexity of the experiment. Using electron tomography, it is therefore far from straightforward to obtain statistically meaningful 3D data, to investigate samples that do not withstand long acquisition, or to perform in situ 3D characterization using this technique. Various acquisition strategies have been proposed to accelerate the tomographic acquisition, and reduce the required electron dose. These methods include tilting the holder continuously while acquiring a projection "movie" and a hybrid, incremental, methodology which combines the benefits of the conventional and continuous technique. In this paper, the different acquisition strategies will be experimentally compared in terms of speed, resolution and electron dose, based on experimental tilt series acquired for various metallic nanoparticles.

1. Introduction

Transmission Electron Microscopy (TEM) is a technique, capable of investigating nano-sized objects down to the atomic scale.[1–4] However, TEM only provides 2D projection images of 3D objects, hereby missing a wealth of information. The interpretation of a 3D structure based on 2D projections alone can indeed be incomplete or unreliable.[5–7] It is therefore clear that a 3D characterization is essential to unravel the structure-property relationship of a given nanomaterial. By combining the powerful imaging capacity of a TEM with the mathematical principles of computed tomography, such investigations are nowadays possible. In 2003, Midgley et al. combined High Angle Annular Dark Field Scanning TEM (HAADF-STEM) with tomography,[5,8] an approach that has been successfully applied to investigate a broad variety of nanostructures.[6,7,9–11] Electron Tomography (ET) experiments are based on acquiring tilt series of projection images over an angular range that is as large as possible to maximize resolution in 3D. To facilitate the ET acquisition, commercial software has been developed, which automates the tilt series acquisition and drives the mechanical rotation of the specimen during the tomography experiment.[12–19] For optimum results, the software implements automatic tracking of the particle of interest between successive tilts by e.g. cross-correlation. Optimization of the defocus can be done by evaluating an image sharpness measure, obtained from projections acquired at different defocus.[20] Nevertheless, even when selecting a low image collection time, the need to perform all of these steps consecutively attributes to a lengthy acquisition. The total time required for the acquisition of a conventional TEM tomography tilt series equals at least 10 minutes. In scanning mode (STEM), only a small fraction of all scattered electrons is captured, due to a restricted detector collection angle, thereby lowering the detection efficiency and prolonging the experimental time to approximately an hour, depending on the complexity of the experiment. Moreover, the pre-acquisition steps (i.e. tracking and focusing) cause additional exposure of the specimen to the electron beam and potentially induce irradiation damage. It is clear that one of the emerging challenges in the field of ET is to improve the speed of tomography experiments, while preserving the quality of the final 3D reconstruction. Specifically, overcoming this challenge is indispensable for enabling:

- ET experiments on beam sensitive materials such as polymers or metal organic frameworks.
- high throughput ET investigations which provide statistically relevant information on for instance particle size or composition.
- *in situ* ET experiments to capture the nanomaterials their 3D dynamic behavior when exposed to operando conditions.

First attempts to accelerate the acquisition process of ET tilt series were made in Bright Field (BF) TEM mode[21–24]. By tilting the holder uninterruptedly while continuously acquiring projection images, the total acquisition time could be reduced to the order of minutes or even less. The total electron dose was lower by at least one order of magnitude when compared to a conventional tilt series. However, BF-TEM tomography is only applicable when investigating non-crystalline, or weakly scattering objects. Otherwise, diffraction effects induce nontrivial BF-TEM contrast for certain orientations of the object with respect to the incoming electron beam, thereby violating the projection requirement for tomographic reconstruction.[5] For such samples, HAADF-STEM is often combined with tomography. In HAADF-STEM diffraction contrast is minimized and mass-thickness contrast predominantly contributes to the image formation. Recently, "continuous" tomography was also introduced for HAADF-STEM imaging.[25] By uninterruptedly tilting the holder and continuously acquiring HAADF-STEM projection images, it became possible to investigate the thermal reshaping and alloying behaviour of metallic nanoparticles under *in situ* conditions.[25–27] However, until now, a thorough experimental evaluation of the accelerated acquisition methodologies with respect to conventional tomography has been lacking. In this paper, we will therefore compare fast HAADF-STEM tomography to conventional HAADF-STEM tomography in a qualitative and quantitative manner, followed by a discussion of the strengths and limitations of both techniques. This paper is structured as follows. First, we introduce the different acquisition methodologies and discuss their advantages and drawbacks. Afterwards, we elaborate on the post-processing involved for fast HAADF-STEM tilt series. In the last section, we present a comparative study of various acquisition methodologies. For each acquisition technique, we will evaluate the speed, the reconstruction quality and the required electron dose.

2. Acquisition methodologies

   2.1 Conventional Tomography

   A conventional ET experiment is composed out of several consecutive steps. To retrieve the 3D structure of a nano-object, a tilt series of images is typically recorded over an angular range of ± (75˚ − 80˚), with a tilt increment of 1–3˚. In this work, a tilt increment of 3˚ was used, since it provides a good compromise between the experimental runtime and the accuracy of the obtained reconstruction when using standard reconstruction algorithms such as the Expectation Maximization (EM) algorithm.[28] At each tilt angle, the sample has to be positioned in the field of view and the image has to be refocused prior to the next acquisition (**Figure 1a**). These consecutive steps contribute to a lengthy acquisition, which may take up to 1 hour or more.

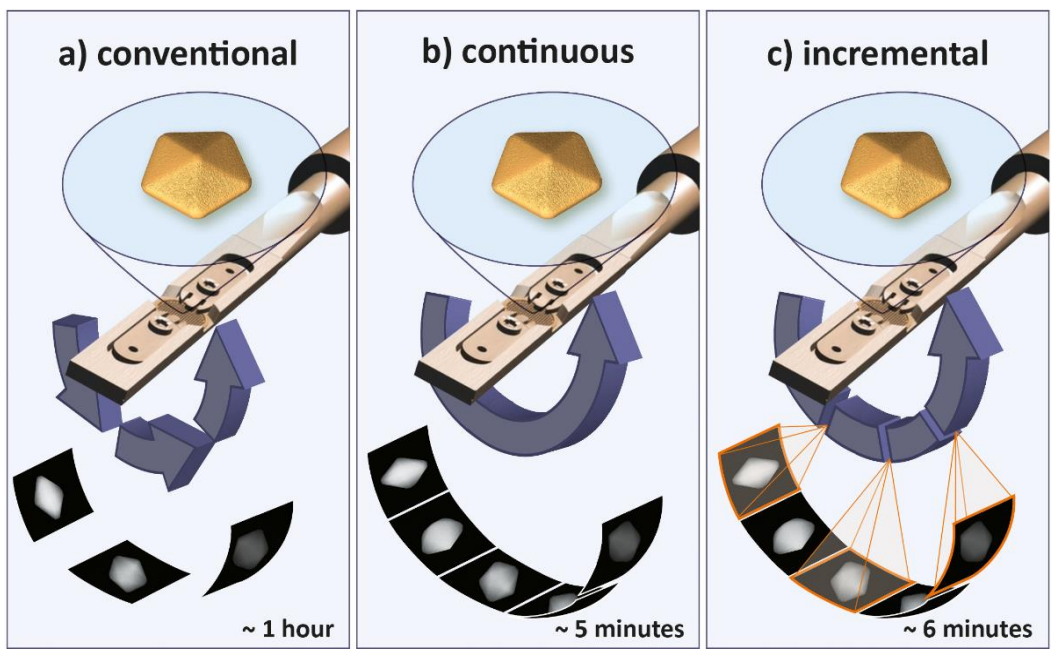

*Figure 1: Graphic illustration of conventional, continuous and incremental tomography. The orange frames in (c) represent stable goniometer positions during the incremental tilting scheme.*

   2.2 Continuous Tomography

   Continuous tomography was proposed to overcome the limitations discussed above, meaning that the holder is rotating uninterruptedly instead of using the classical step-by-step procedure, and intermediate refocusing and repositioning is performed manually at the same time.[25] In this manner, a projection "movie" is recorded rather than a set of projection images (**Figure 1b**). By avoiding the individual repositioning and refocus steps it becomes possible to reduce the total acquisition time for a tilt series by at least a factor of 10. The effective reduction in experimental runtime depends on the mechanical stability of the microscope stage, the sensitivity of the detector and the detector read-out time. The related electron dose reduction will be further discussed in Section 5 (**Table 3**).

As can be expected, the time gain comes at a cost of the image quality. For example, because of the continuous tilt, the nano-object rotates around the tilt axis while acquiring a single HAADF-STEM "frame". Therefore, it is essential to keep the tilt angle per frame small enough to treat the frames as static projection images. In this work, we selected a frame time of 1 s and a tilt speed such that the sample was tilted from -75˚ to 75˚ in a total time of 5 minutes. This corresponds to a tilt increment of 0.6° per image, which is sufficiently small to treat the continuous projection movie as being built up out of static projection images, whereas larger tilt angles per frame may lead to blurring effects that deteriorate the reconstruction. An additional factor that restricts the tilting speed is the capability of the microscope operator to keep the nano-object in the field of view during the tilt, which in turn depends on the mechanical stability of the microscope stage and magnification used. Selection of the frame time is also defined by this factor, since for longer frame times at a similar tilting speed, significant drift of the nano-object might occur over the course of the acquisition of a single HAADF-STEM frame. As a result, motion artefacts appear which deteriorate the quality of the continuous acquisition. On the other hand, much shorter frame times lower the signal-to-noise ratio of the recorded images, which impedes the manual tracking and refocusing of the particle. Therefore, a compromise has to be made between the tilting speed and frame time. An example of a raw continuous tilt series is provided in the **Supplementary Movie 1.**

2.3 Incremental Tomography

A major bottleneck for continuous tomography is that the goniometer stage and tomography holders of common electron microscopes are not originally designed for fast, continuous, tilting. As a result, high frequency vibrations arise when tilting at high speed. Such high frequency vibrations give rise to jitter-like artefacts in the HAADF-STEM projection images, as illustrated in **Figure 2a.** Whereas the influence of the vibrations can be minimized through post-processing methods, as will be discussed later on, it is fundamentally preferential to avoid such noise during the experiment. Therefore, a hybrid acquisition strategy was proposed that combines the advantages of the conventional and continuous acquisition method, referred to as the "fast incremental tomography" (**Figure 1c**). During the incremental approach, tilting is performed in a step-wise manner, similar to the conventional method. At each angle, a certain relaxation time is imposed which can be used to reposition and refocus the sample manually. The projection images however are acquired continuously, again resulting in a projection movie. Given the incremental tilting and the imposed relaxation time, projection frames are also obtained while the goniometer is in a stable state (**Figure 1c,** orange framed projections), thereby efficiently removing the source of the high frequency vibrations. As a result, the incremental approach provides a compromise between the speed of a continuous acquisition, and the image quality of a conventional acquisition.

The dominant time gain for incremental tomography results from the acceleration of the pre-acquisition steps. Indeed, while commercially available software typically requires multiple images to evaluate whether the particle of interest is in the field of view and in focus, an experienced TEM operator can perform such actions much more rapidly. Here, we selected a frame time of 1 s and a tilt increment of 2° for the incremental acquisition. At every angular position, a relaxation period of 4 s was imposed, which could be used to reposition and refocus on the particle of interest and provided sufficient time for the goniometer to stabilize. This leads to a total acquisition time of 6 minutes. The stage tilt for the incremental approach was programmed through the scripting interface available in Thermo Fisher Scientific microscopes using the Python programming language.

3. Pre-processing methodologies

Prior to the experimental evaluation of three different acquisition strategies in terms of resolution and electron dose, we discuss all computational pre-processing steps involved in handling the fast tilt series. The different steps will be explained using a continuous tilt series of a Au nano decahedron.

*3.1 Noise correction*

**Figure 2a** shows a HAADF-STEM projection frame which was extracted from continuous tilt series of a Au nano decahedron. Jitter along the fast scanning axis (white arrow) can be observed due to mechanical instabilities that occur during the continuous acquisition.[24] Because of the directionality and broad frequency distribution of the jitter, the artefacts appear as a stripe in the Fourier transform of the acquired images (**Figure 2b**, white arrow). The first step of data pre-processing therefore consists of applying a directed low pass filter in Fourier space (**Figure 2c**), resulting in an improved image quality (**Figure 2d**). The width, orientation, inner and outer radius of the filter are to be manually tuned, since they depend on the acquisition parameters and the structure of the studied object. Indeed, increasing the width of the filter improves the denoising, but introduces additional directional blurring as well. Consequently, a compromise has to be made.

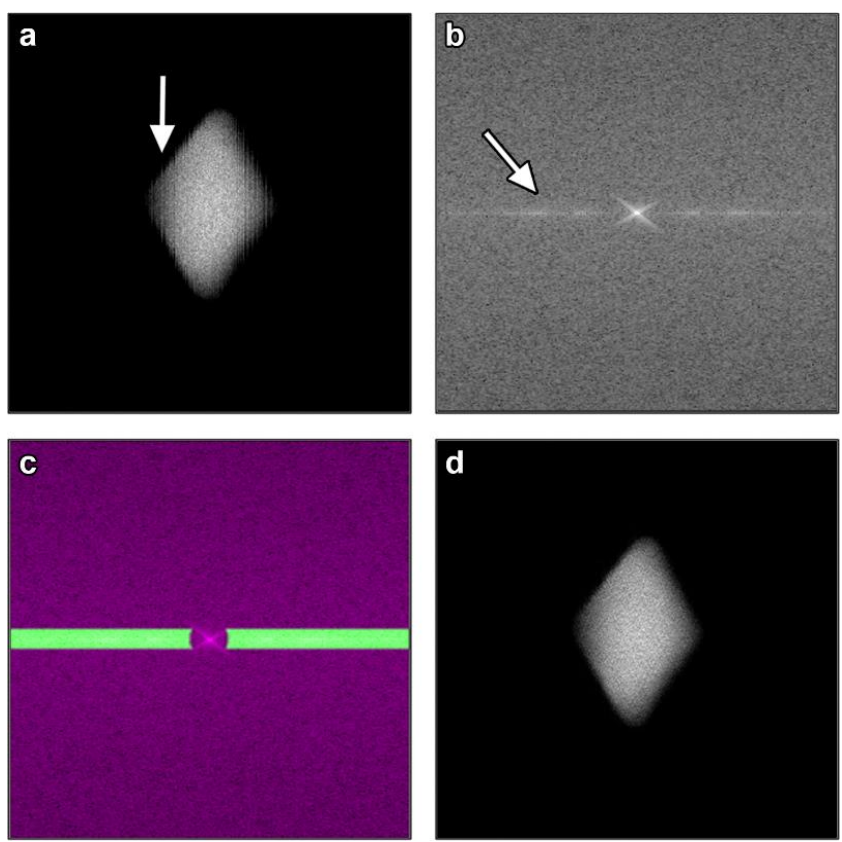

*Figure 2*: *(a) HAADF-STEM image obtained during the fast continuous acquisition of a single nanoparticle, revealing distortions along the fast scanning axis. (b) The amplitude of the Fourier transform of (a). (c) The directional low pass filter used to remove these frequencies and restore the projection image (d).*

*3.2 Angle assignment*

In order to use the frames, extracted from the continuous series, as input for a tomographic reconstruction, an angle needs to be assigned to each projection. Given our static projection approximation, we assigned a constant angle value to each projection image. The constant angle value was determined as the average tilt angle of the pixels in that image, as illustrated in **Figure S1**.

*3.3 Alignment*

The projection images are aligned with respect to each other using a cross-correlation approach. For the incremental acquisition strategy, pre-processing starts with the alignment since the object is kept stable between consecutive tilts. The angle assignment for the incremental acquisition is therefore more convenient as it can be determined while the goniometer stage is in a stable position (i.e. not rotating). In addition, the application of a directed low pass filter is no longer required since the high frequency vibrations are already experimentally prevented. However, for both fast acquisition techniques, the mechanical instability of the specimen holder may cause considerable drift. As a result, the sample can partially move out of the field of view and/or movement artefacts can occur. Hereby, the standard cross-correlation approach, where each image is aligned to the previous one, performs sub-optimally. In **Figure S2a**, the average of a fast continuous tilt series is shown after it was aligned based on only the previous projection image. It can be seen that no precise alignment is obtained. Therefore, all projection images are aligned to the average of a user selected number of previous projection images (typically 5 to 10). As can be seen from **Figure S2b**, this considerably improves the stability of the alignment procedure. Furthermore, it reduces the negative influence of defocused and/or deformed projection images, caused by manual tracking and focusing during the fast acquisition, present in the tilt series. After performing the image alignment, the tilt axis alignment is carried out using the conventional approach.[30]

*3.4 Image Selection and Reconstruction*

Distorted projection images do not only hamper the alignment, they can also negatively influence the reconstruction accuracy. After the initial alignment, deteriorated projection images are therefore removed from the tilt series in an automated manner (**Figure 3a**). First, an initial reconstruction is computed making use of the ASTRA toolbox implementation of the EM reconstruction algorithm[31–33]. Next, forward projections are calculated for all angles present in the fast tilt series. Comparison between the forward projections of an intermediate reconstruction and experimentally acquired projection images allows to determine which projection images have the least quality. **Figures 3b-c** depict one of the removed projection images together with their forward projected counterpart. Sample drift over the course of the acquisition of a single HAADF-STEM frame results in motion artefacts, which can be recognized as skewing of the object of interest (**b**). Such artefacts are not present in the forward projection of the intermediate reconstruction (**c**), since only consistent features coming from the real structure of object are promoted by the 3D reconstruction algorithm.

To assess which projection images predominantly deteriorate the reconstruction accuracy, we evaluate the absolute deviation around the median of the normalized root mean square error (RMSE) between the experimental and calculated projection images[34]. The absolute deviation of the median provides a measure for the relative difference of the RMSE of a single projection image to the median of the RMSE of all projection images. This approach is therefore an efficient method to detect outliers (i.e. the most deteriorated projection images) within the tilt series. We hereby use the following formula:

$$RMSE_i = \sqrt{\sum_j^N \left(p_j^{exp,i} - p_j^{calc,i}\right)^2 / N}. \quad [1]$$

The experimental continuous projection images and their calculated forward projected counterparts are respectively defined as $p^{exp}$ and $p^{calc}$. $N$ corresponds to the number of pixels in the projection images. Once the error measure is obtained for each projection image, the absolute deviation around the median is calculated as follows:

$$RMSE_i' = \frac{|RMSE_i - median(RMSE)|}{median(RMSE)}. \quad [2]$$

In **Figure 3d**, the RMSE graph for an experimental tilt series is plotted (blue line). Images acquired at high tilt are generally of lower quality because of detector shadowing by the holder and/or grid and consequently they intrinsically differ more from their calculated counterparts. Therefore, a low order polynomial trend was first removed from the RMSE graph. By manually thresholding (red line) the graph, the most distorted images can be detected and consequently removed from the tilt series. Finally, the reduced projection stack is realigned, as discussed above, and this procedure is iteratively repeated until the values for the RMSE′ are below the preset threshold. Given the large amount of projection images in a fast tilt series, in comparison to a conventional tilt series, removal of the distorted projection images will not lower the quality of the reconstruction. Finally, in this work, the 3D reconstruction is calculated by using the reduced tilt series as input to for the EM reconstruction algorithm for 20 iterations.[28]

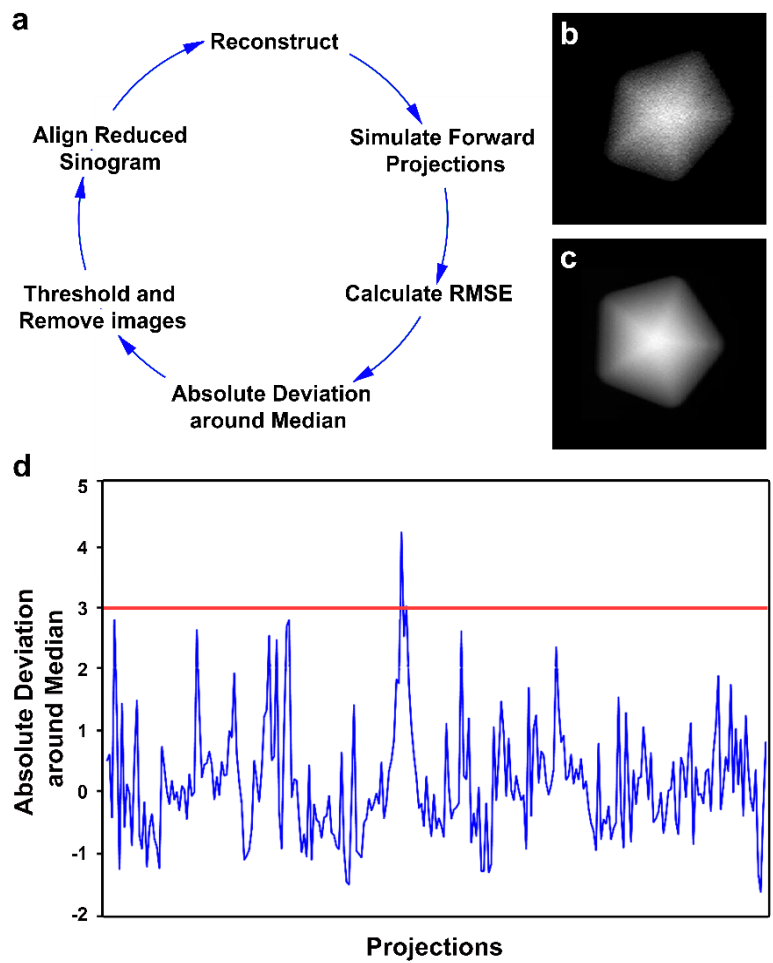

*Figure 3*: *(a) Illustration of the iterative scheme, used to remove deteriorated projection images. (b) Experimental image and (c) calculated projection of an intermediate reconstruction of the object. (d) Error metric for images in an experimental tilt series and the threshold used to exclude corrupted images.*

4. Results
   4.1 Resolution

To evaluate the difference in reconstruction quality between the conventional, continuous and incremental approach, we acquired three HAADF-STEM tilt series of the same nanoparticle, using the three different acquisition methodologies, for five different sample morphologies. For this purpose, we selected Au and Au@Ag nanoparticles, since they are relatively resistant to the electron beam. Each conventional series consisted out of 51 projection images, acquired over an angular range of ± 75° with a tilt increment of 3° using either a Thermo Fisher Scientific Titan TEM operated at 300 kV or a Thermo Fisher Scientific Tecnai Osiris TEM operated at 200 kV. For every conventional projection image, a frame time of 6 s was used. Additional focus and repositioning steps lead to a total acquisition time of approximately 1 hour per conventional series. Next, fast continuous and incremental tilt series were collected from the same nanoparticles, using the same microscopes, with identical magnification and angular range. For the continuous acquisition, a frame time of 1 s was selected and the goniometer rotation speed was set such that the sample was tilted from -75° to 75° in 5 minutes of time. In this manner, a good compromise between signal-to-noise ratio and experimental runtime was obtained. For the incremental acquisition, a frame time of 1 s was selected as well, in combination with a tilt increment of 2° and a relaxation period of 4 s. This results in a total experimental runtime of 6 minutes. Details on the experimental parameters of fast series are listed in **Table 1**. The conventional series were acquired under similar experimental conditions, apart from the longer dwell- and frame time and therefore with a higher electron dose. The electron dose and experimental runtime will be further discussed in Section 5 (**Table 3**).

|  | *Sample description* | *Frame size (px)* | *Frame time (s)* | *Pixel size (pm)* | *Dwell time (µs)* | *Electron dose per image (e/Å$^2$)* | *Microscope* |
|---|---|---|---|---|---|---|---|
| *Sample 1* | Au octopod | 512 x 512 | 1 | 1086 | 3 | 2.65 x 10$^6$ | Titan (300 kV) |
| *Sample 2* | Au octopod | 512 x 512 | 1 | 1193 | 3 | 2.20 x 10$^6$ | Osiris (200 kV) |
| *Sample 3* | Au@Ag nanorod | 512 x 512 | 1 | 771 | 3 | 5.26 x 10$^6$ | Titan (300 kV) |
| *Sample 4* | Au@Ag nanotriangle | 512 x 512 | 1 | 597 | 3 | 8.77 x 10$^6$ | Osiris (200 kV) |
| *Sample 5* | Au nanostar | 512 x 512 | 1 | 388 | 3 | 2.08 x 10$^7$ | Titan (300 kV) |

***Table 1:*** *Experimental details of the tilt series acquired for evaluating the different acquisition methodologies.*

Reconstructions were calculated, based on the continuous and incremental tilt series and compared to the reconstruction obtained from the conventional tilt series. Both the continuous and incremental series were processed as previously discussed in Section 3. In **Figure 3**, a qualitative comparison is made between all acquisition techniques for a Au octopod which was investigated using the Thermo Fisher Scientific Titan TEM operated at 300 kV (sample 1). At the top row of the figure, a 3D rendering of the conventional reconstruction and two central orthoslices through the reconstruction are presented. The orientation of the orthoslices is indicated in the top right corner. At the second row, similar results are provided for the reconstruction obtained from a continuous tilt series, whereas the bottom row displays the results for the incremental approach. In **Figure S3,** a similar comparison is made for the Au octopod which was investigated using the Thermo Fisher Scientific Tecnai Osiris TEM operated at 200 kV (sample 2). In both cases, little difference is observed between the different reconstructions. The continuous approach causes a slight blurring which can either result from the application of the low pass filter, the continuous tilting, or the remaining deformed and/or defocused projection images still present in the tilt series. Such blurring does not appear to be present when using the incremental acquisition strategy. The Au octopods (sample 1 & 2) are relatively large and the acquisitions were therefore performed at moderate magnifications with a pixel size of approximately 1000 pm (**Table** 1). Since the vibrations caused by the mechanical instabilities occur at a scale which is at least one order of magnitude smaller, their effect is almost negligible for such moderate magnifications. However, because of the octopods their relatively large size, cupping artefacts (i.e. the underestimated intensity in the interior) can be seen in both reconstructions.[35] In **Figure 4** and **Figure S4**, reconstructions of Au@Ag nanoparticles (samples 3 & 4) are compared. To accurately investigate the interface between Au and Ag, a smaller pixel size was used (**Table** 1). Blurring induced by the continuous acquisition prevents one from observing a clear separation between both phases. However, the incremental reconstruction considerably reduces the amount of blurring. Thereby enabling an accurate investigation of the Au-Ag interface and potential intermixture of both elements.[27] For the anisotropic nanostar[36] (sample 5, **Figure 5**), the amount of blurring becomes more substantial and conceals small features present in the reconstruction. This is likely related to the increased complexity of morphology of the NSs and their small-scale features. We expect that the remaining artefacts present in the projection images have a stronger influence on the reconstruction of nanostructures with a complex 3D shape, in comparison to reconstructions of more simple geometries. The incremental acquisition however, reduces the mechanical instabilities and consequently minimizes the amount of blurring.

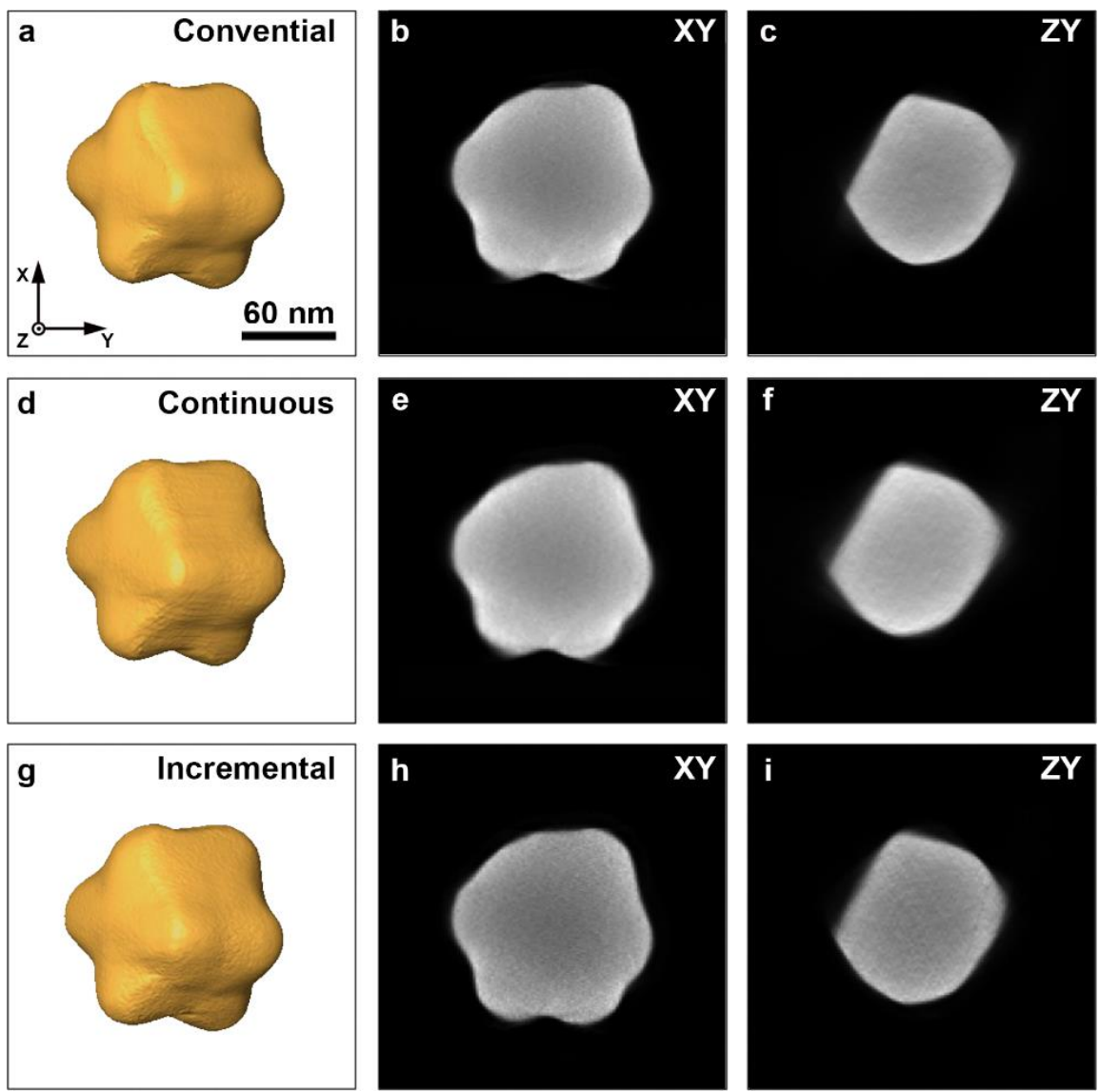

*Figure 3*: *(a-c) 3D rendering of the reconstruction of the conventionally acquired tilt series of a Au octopod and two central orthoslices through the reconstruction. (d-f) 3D rendering of the reconstruction of the continuously acquired tilt series and 2 central orthoslices through it. (g-i) 3D rendering of the reconstruction of the incrementally acquired tilt series and 2 central orthoslices through the reconstruction.*

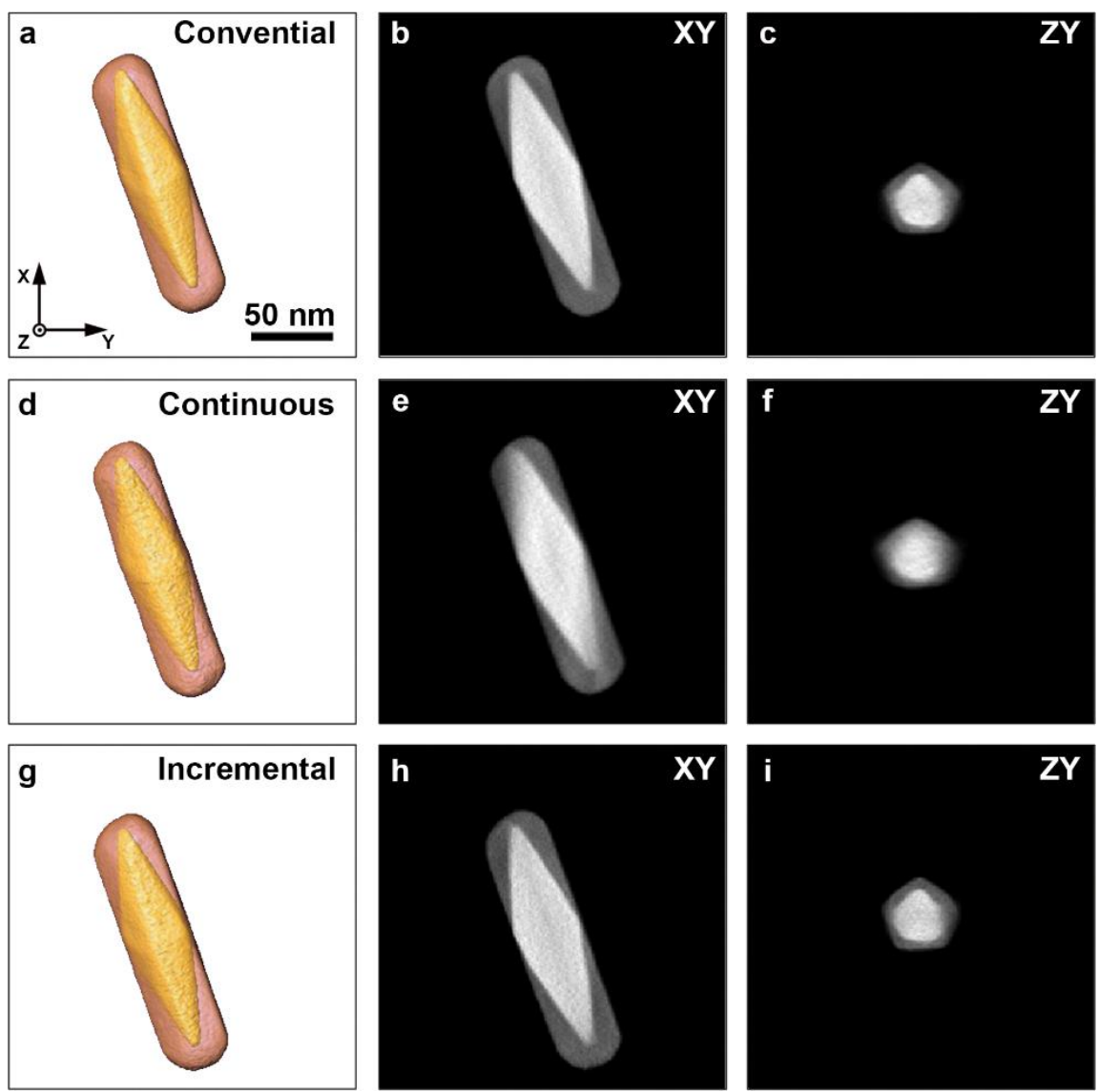

***Figure 4***: *(a-c) 3D rendering of the reconstruction of the conventionally acquired tilt series of a Au@Ag nanorod and two central orthoslices through the reconstruction. (d-f) 3D rendering of the reconstruction of the continuously acquired tilt series and 2 central orthoslices through it. (g-i) 3D rendering of the reconstruction of the incrementally acquired tilt series and 2 central orthoslices through the reconstruction.*

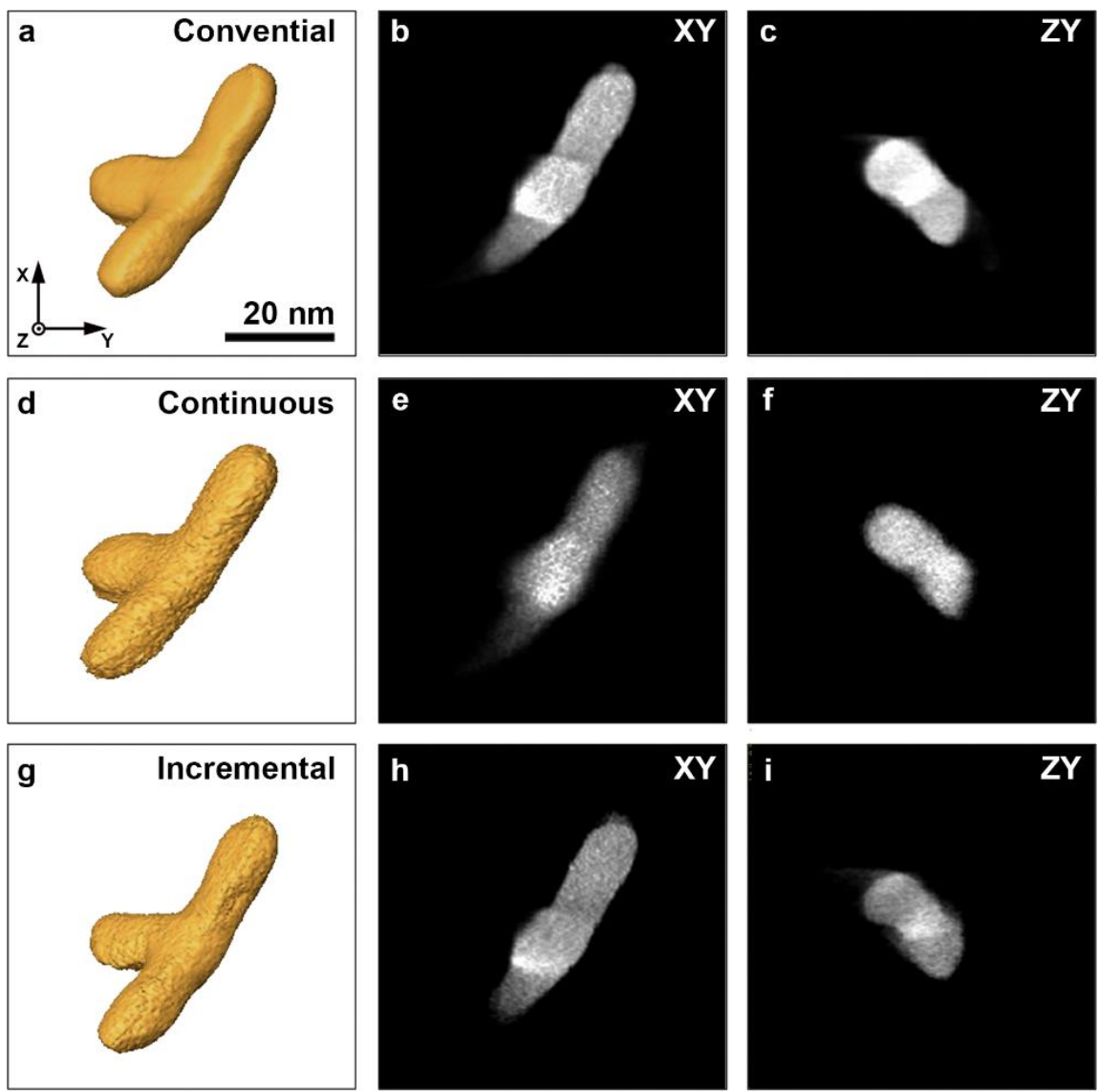

*Figure 5*: *(a-c) 3D rendering of the reconstruction of the conventionally acquired tilt series of a Au nanostar and two central orthoslices through the reconstruction. (d-f) 3D rendering of the reconstruction of the continuously acquired tilt series and 2 central orthoslices through it. (g-i) 3D rendering of the reconstruction of the incrementally acquired tilt series and 2 central orthoslices through the reconstruction.*

To investigate the differences in a quantitative manner, the shape- and volume error were calculated for the reconstructions obtained from the fast tilt series. The shape error $E_s$ was calculated as a weighted sum over absolute differences for each reconstruction with respect to the reference (i.e. the reconstruction obtained from the conventional series) as follows:

$$E_s = \frac{\sum \left| Rec_{fast}^{seg} - Rec_{conv}^{seg} \right|}{\sum Rec_{conv}^{seg}} \times 100 \quad [3]$$

$Rec_{conv}^{seg}$ and $Rec_{fast}^{seg}$ refer to the segmented 3D reconstructions of the conventional and fast (i.e. continuous or incremental) series respectively. The sum over $Rec_{conv}^{seg}$ defines the volume of the reference reconstruction. In this manner, the shape error determines the relative difference in shape between the reconstructions based on the reference series and the series acquired using the fast acquisition schemes, as a percentage of the total amount of misclassified voxels. The threshold for segmentation was determined by Otsu's method[37] for each reconstruction separately. Before calculating the absolute difference, both reconstructions were optimally aligned using a MATLAB implementation of intensity-based rigid registration. Through gradient descent optimization, this operation minimizes the Mean Square Error (MSE) between the two volumes through translation and/or rotation. In addition, the volume error ($E_V$), which corresponds to the relative volume difference between the different reconstructions, was determined. The outcome of this quantitative assessment is summarized in **Table 2**. For the Au@Ag nanoparticles, error measures were calculated for each separate phase. **Table 2** shows that the incremental approach improves the reconstruction accuracy considerably. Especially for the Au@Ag nanoparticles a substantial reduction of both errors is found. It can be observed that the shape error for the Au octopod investigated using a Titan TEM (sample 1) is lower that the shape error for the near identical octopod investigated using a Tecnai Osiris TEM (sample 2), independent of the used acquisition approach. This demonstrates that the used microscope strongly impacts the accuracy that can be achieved by the fast acquisition methodologies. The improved stability of the goniometric stage of a Titan TEM in comparison to a Tecnai Osiris TEM, is likely responsible for the observed differences. Clearly, as the stability of the goniometer stage design improves, the fast approaches will produce even more accurate results. We furthermore found that magnification plays a vital role as well. Indeed, for the Au octopod (sample 1) investigated using the Titan microscope and a pixel size of 1086 pm, a shape error of 2.95 % was found for the continuous acquisition. The nanostar (sample 5) which was investigated using the same instrument but using a much smaller pixel size (338 pm), resulted in a significantly higher shape error, equal to 16.30 %. This confirms that at such small scales, the vibrations induced by the mechanical instabilities have a more detrimental effect on the reconstruction accuracy.

|  |  | $E_S$ continuous | $E_S$ incremental | $E_V$ continuous | $E_V$ incremental |
|---|---|---|---|---|---|
| *Sample 1* | Au octopod (Titan) | 2.95 % | **2.26 %** | 2.05 % | **1.69 %** |
| *Sample 2* | Au octopod (Osiris) | 6.24 % | **4.70 %** | 5.25 % | **2.74 %** |
| *Sample 3, Au* | Au@Ag rod (Titan) | 8.09 % | **3.77 %** | 5.61 % | **0.25 %** |
| *Sample 3, Ag* |  | 14.96 % | **8.34 %** | 4.08 % | **0.42 %** |
| *Sample 4, Au* | Au@Ag nanotriangle (Osiris) | 9.15 % | **8.81 %** | 1.37 % | 1.78 % |
| *Sample 4, Ag* |  | 24.00 % | **16.24 %** | 16.85 % | **4.68 %** |
| *Sample 5* | Au nanostar (Titan) | 16.30 % | **12.90 %** | 5.88 % | **1.23 %** |

*Table 2*: The shape- and volume errors calculated between the reconstructions obtained from the continuous and incremental acquisition strategies for the different nanoparticles.

4.2 Electron Dose

In addition to the benefit in experimental runtime, the fast acquisition approach drastically reduces the total accumulated dose. As introduced earlier, a conventional tomography experiment is composed out of a series of steps which are performed in a sequential manner:

- Tilt the object to the next angular position.
- Track the particle of interest.
- Optimize the focus.
- Acquire and save the respective image.

Commercial software has been developed to automate the acquisition procedure.[12–19] While this facilitates the execution of the tomography experiment, it also requires additional electron dose. In **Table 3**, the calculated electron doses are presented, based on typical experimental parameters for the various acquisition strategies. Considering an electron beam with a 50 pA beam current, the electron dose was calculated as:

$$Dose = \frac{beam\ current\ x\ frame\ time}{electron\ charge\ x\ (pixel\ size)^2} \ x\ number\ of\ frames \qquad [3]$$

For the electron dose calculations, we use similar experimental parameters as before and consider that the conventional tilt series were acquired over an angular range of ± 75° with a tilt increment of 3° and a frame time of 6 s. In addition, the tracking and refocussing needs to be taken into account, since these operations require additional illumination of the sample and therefore contribute to the accumulated electron dose. To accurately track the particle of interest, an additional projection image needs to be acquired after the tilt. This image is typically acquired at a lower magnification, to make sure that the particle of interest does not move out of field of view over the course of tilting the stage. To decrease the required electron dose and accelerate the experimental runtime, the tracking is conventionally performed with a lower frame time. For the electron dose calculations, we therefore assumed a four times larger pixel size for the tracking, and a short frame time of 3 s, as listed in **Table 3**. The refocus step is done by acquiring a through focus series. Conventionally, 7 images are acquired at different defocus values. By analysing, for instance, the variance of the images, the best defocus can be determined and used for the next acquisition.[20] Also here, a lower frame time and/or frame size is used to limit the influence of the electron beam. For the electron dose calculations, we therefore considered a frame time and size of respectively 3 s and 256 x 256 pixels.

For both fast acquisition techniques, no additional electrons are required for the tracking and refocussing since these corrections are simultaneously performed. When comparing the total calculated electron dose for the different acquisition strategies (**Table 3**), it is clear that the required focus step for the conventional methodology leads to a much higher dose. For the selected experimental parameters, both the continuous and incremental acquisition require approximately only 50 – 60% of the dose necessary for a conventional experiment. This demonstrates that the fast acquisition methodologies are not only useful to increase the throughput of ET, but may also allow the 3D investigation of nanomaterials which are sensitive to the electron beam. It must be stressed that the dose for a conventional experiment is an optimistic estimate since the automated procedures may fail to properly focus or reposition the particle of interest. This may may occur when multiple nanostructures are in field of view or when the material under investigation is composed of low scattering elements. If the automated software fails, a longer experimental runtime and therefore a higher electron dose is required.

|  | **Conventional** | **Continuous** | **Incremental** |
|---|---|---|---|
| **Acquisition** | Beam current: 50 pA<br>Dwell time: 19 μs<br>Frame time: 6 s<br>Frame size: 512 x 512<br>Pixel size: 500 pm<br>Tilt range: ± 75°<br>Tilt increment: 3°<br># images: 51 | Beam current: 50 pA<br>Dwell time: 3 μs<br>Frame time: 1 s<br>Frame size: 512 x 512<br>Pixel size: 500 pm<br>Tilt range: ± 75°<br># images: 300 | Beam current: 50 pA<br>Dwell time: 3 μs<br>Frame time: 1 s<br>Frame size: 512 x 512<br>Pixel size: 500 pm<br>Tilt range: ± 75°<br>Tilt increment: 2°<br>Relaxation time: 4 s<br># images: 380 |
| Dose (e/Å$^2$) | 3.83 x 10$^9$ | 3.75 x 10$^9$ | 4.75 x 10$^9$ |
| **Tracking** | Beam current: 50 pA<br>Dwell time: 9 μs<br>Frame time: 3 s<br>Frame size: 512 x 512<br>Pixel size: 2000 pm<br># images per tilt: 1 | | |
| Dose (e/Å$^2$) | 1.2 x 10$^8$ | | |
| **Focus** | Beam current: 50 pA<br>Dwell time: 9 μs<br>Frame time: 3 s<br>Frame size: 256 x 256<br>Pixel size: 1000 pm<br># images per tilt: 7 | | |
| Dose (e/Å$^2$) | 3.35 x 10$^9$ | | |
| **Total Dose (e/Å$^2$)** | **7.29 x 10$^9$** | **3.75 x 10$^9$** | **4.75 x 10$^9$** |

*Table 3*: *Dose calculation for the various acquisition strategies.*

5. Discussion

In this work, we presented two methodologies to perform fast ET in HAADF-STEM mode, which enable one to greatly reduce the experimental runtime for ET. We discussed all techniques in terms of acquisition time, reconstruction quality and electron dose. Our findings are summarised in **Table 4**. We show that by using dedicated processing methodology, there is a limited difference in reconstruction accuracy for large and isotropic nanoparticles. Even more so, such results can be obtained in only a fraction of the time necessary to perform a conventional experiment. For relatively small anisotropic nanoparticles with a more complex shape, the discrepancy increases but the overall 3D shape remains comparable. It was demonstrated that the proposed incremental methodology considerably increases the reconstruction accuracy, in comparison to the continuous method, especially for anisotropic nanoparticles.

In terms of electron dose, the necessary pre-acquisition steps for a conventional acquisition, refocussing in particular, attribute to a significantly higher accumulated electron dose. Although the frame time and/or frame size can be decreased to minimize their contribution, this would compromise the retrieval of the most optimal defocus. By replacing the fully automated workflow with a semiautomatic one, ET experiments can be optimized for speed and electron dose. Both fast acquisition approaches demonstrate that an experienced TEM operator can perform the pre-acquisition actions in a much more efficient manner, requiring less electron dose, in a guided workflow as proposed in this paper. This is further exemplified by the resulting electron dose reduction of approximately 50%. Such results are of high interest for the studies of radiation sensitive materials which are currently still extremely challenging since the samples will degrade and deform when exposed to relatively high electron doses.

Although incremental tomography leads to a higher the reconstruction accuracy compared to continuous tomography, it can be experimentally more challenging. During the incremental acquisition, the sample is tilted in a step-wise manner between consecutive angels. In order to minimize the experimental runtime, these tilt intervals are performed at such a high speed that it is difficult to manually track the nano-object of interest at the same time. As a consequence, the nano-object more easily drifts out of the field of view during the acquisition. It is therefore clear that for incremental tomography, the experimental runtime is limited by the chosen tilt increment and relaxation time. Large tilt increments and short relaxation periods will speed up the acquisition, but complicate the tracking. We expect that development of more stable goniometric stages for (S)TEM instruments will bring a significant improvement in the quality of the data provided by the incremental technique. Currently, however, continuous tomography holds most promise in terms of experimental runtime. This is evidenced by the decrease in experimental runtime from approximately 1 hour to 5 minutes. Nevertheless, for *in situ* tomography experiments where dynamical changes occur within seconds, the total acquisition time has to be further lowered by at least a factor of 10. However, when tilting even faster, high acquisition rate detectors and electron beam deflectors are required to ensure the static projection approximation or dedicated algorithms need to be developed which take the in-frame rotation into account.[22,23,38] In addition, the manual tracking and refocussing will become increasingly challenging, and should be ideally replaced by optimized automatic routines.

From our results, it is clear that for each ET experiment a careful choice needs to be made concerning the used microscope, experimental settings and acquisition strategy. As a general guideline, continuous tomography is to be used when acquisition time is priority and uninterrupted feedback is required for tracking the nano-object during the acquisition. This is for instance the case when the goniometric stage has a limited stability or when a relatively high magnification is required. However, if there is little sample drift during the tilt, incremental tomography is preferential, since it improves the reconstruction accuracy, especially for small features of interest. Conventional tomography remains, for the time being, the ideal technique when reconstruction quality and resolution are most important and a long experimental runtime and a high electron dose can be tolerated.

|  | **Conventional (manual)** | **Conventional (software)** | **Continuous** | **Incremental** |
|---|---|---|---|---|
| **Reconstruction quality** | ++ | ++ | – | + |
| **Acquisition Time** | – – | – | ++ | + |
| **Electron Dose** | – – | – | ++ | + |

***Table 4**: Summarised comparison of the various acquisition strategies.*

6. Conclusions

In conclusion, we compared 3 different acquisition strategies for HAADF-STEM tomography series; conventional, continuous and incremental. We discussed the required pre-processing steps, necessary to obtain a reliable reconstruction based on the fast continuous and fast incremental series. It was demonstrated that the fast approaches show great promise for accelerating the tomographic experiment by approximately a factor of 10, and lowering the electron dose by a factor of 2. We observed for different nanoparticles that the general morphology of the 3D reconstruction is conserved quite well when using a fast acquisition methodology. We can therefore still obtain valuable information on the nano-objects investigated using the fast approaches that only require a fraction of the time and electron dose necessary to perform a conventional experiment. The benefits of the findings can be applied to enable tomography on beam sensitive materials, collecting more statistical relevant 3D data on nanostructures or improve the temporal resolution of 3D imaging for *in situ* studies.

## 8. Acknowledgements

We acknowledge Prof. Luis M. Liz-Marzán and co-workers of the Bionanoplasmonics Laboratory, CIC biomaGUNE, Spain for providing the Au@Ag nanoparticles, Prof. Sara. E. Skrabalak and co-workers of Indiana University, United States for the provision of the Au octopods and Prof. Teri W. Odom of Northwestern University, United States for the provision of the Au nanostars. H.V. acknowledges financial support by the Research Foundation Flanders (FWO grant 1S32617N). S.B acknowledges financial support by the Research Foundation Flanders (FWO grant G.0381.16N). This project received funding as well from the European Union's Horizon 2020 research and innovation program under grant agreement No 731019 (EUSMI) and No 815128 (REALNANO). The authors acknowledge the entire EMAT technical staff for their support.

## 9. Supplementary Information

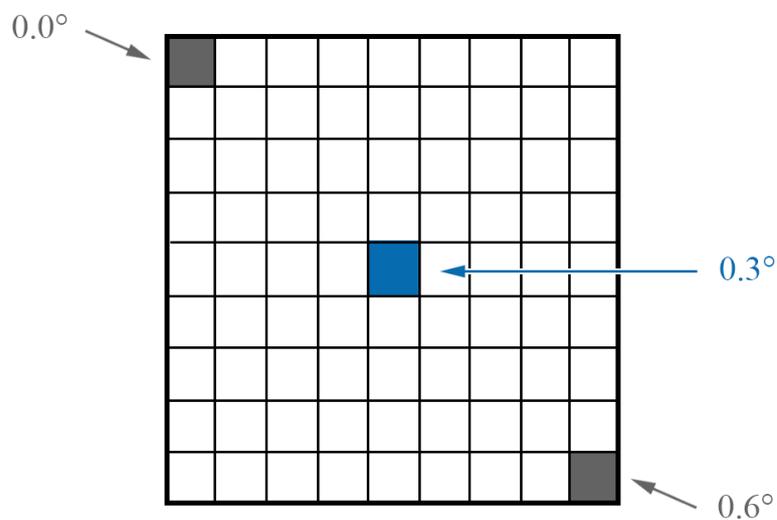

*Figure S1: For the static approximation, each projection frame was assigned its average tilt angle.*

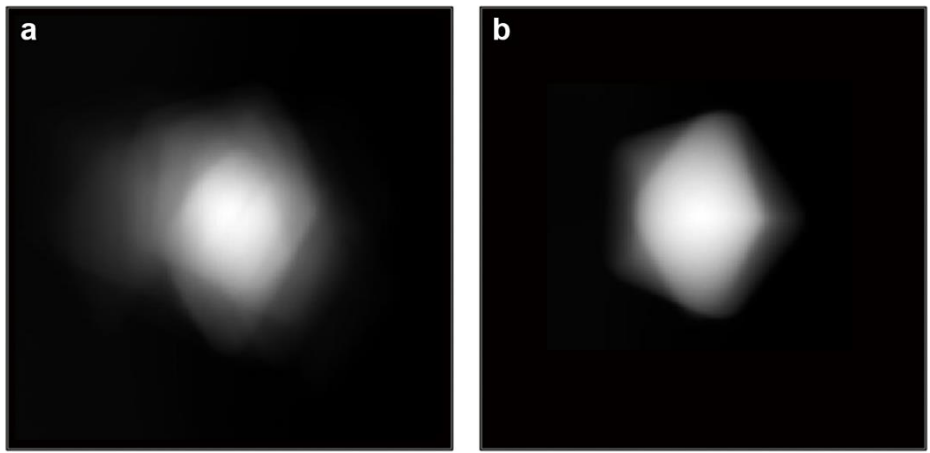

*Figure S2: Comparison of the average of the aligned tilt series for which the alignment was performed using (a) only the previous projection image or (b) the average of the 5 previous projection images.*

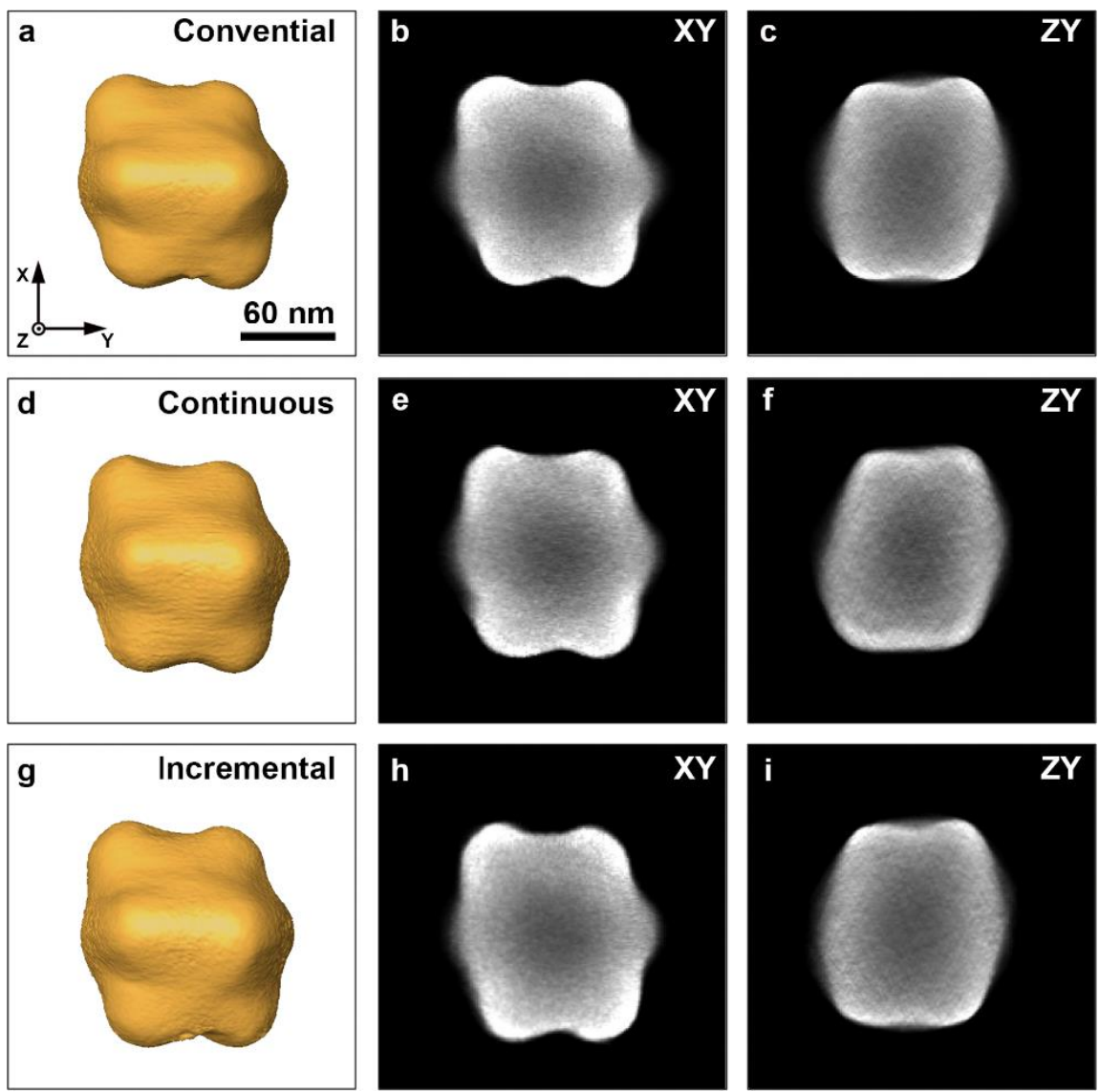

***Figure S3***: *(a-c) 3D rendering of the reconstruction of the conventionally acquired tilt series of a Au octopod and two central orthoslices through the reconstruction. (d-f) 3D rendering of the reconstruction of the continuously acquired tilt series and 2 central orthoslices through it. (g-i) 3D rendering of the reconstruction of the incrementally acquired tilt series and 2 central orthoslices through the reconstruction.*

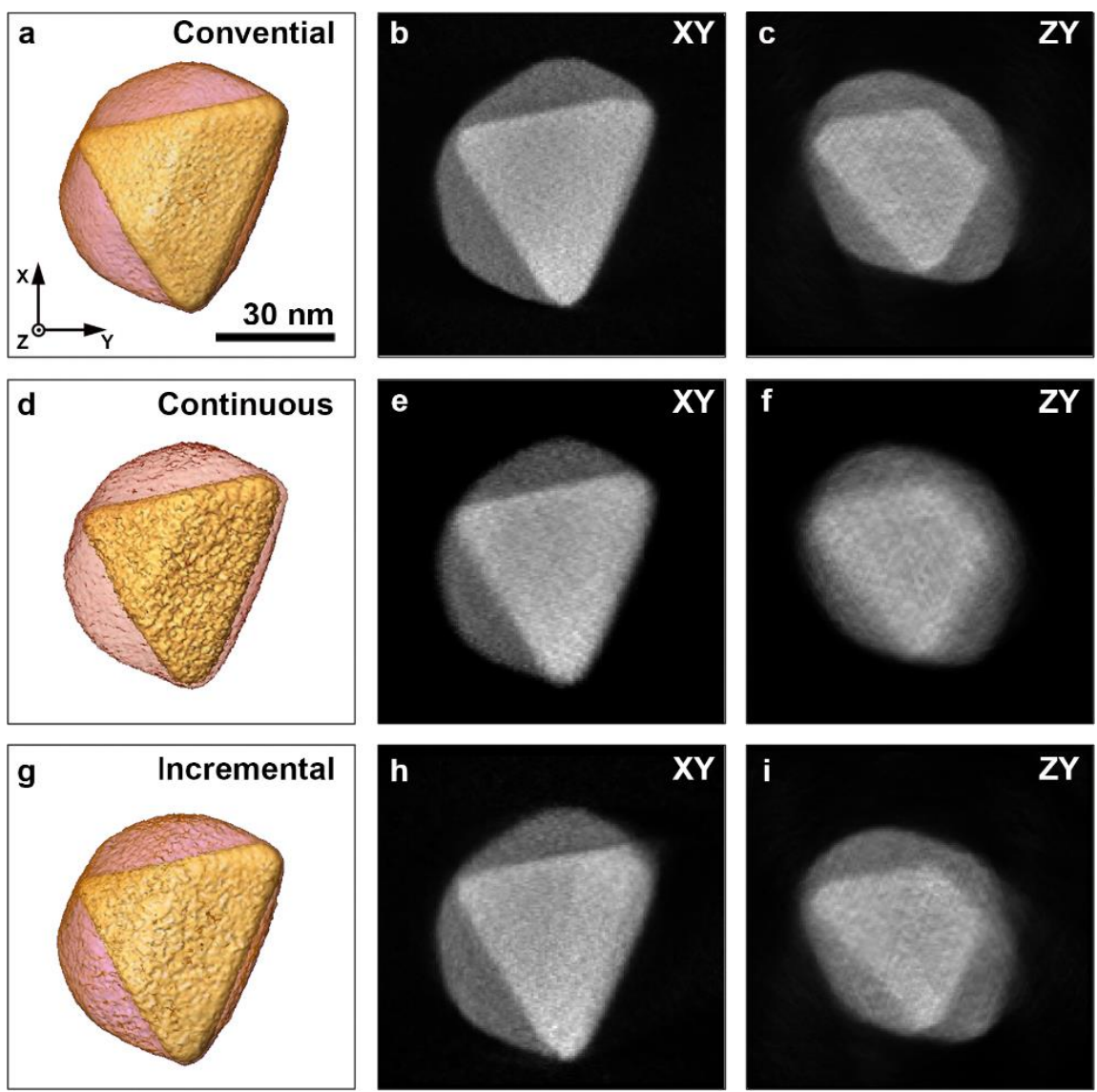

*Figure S4*: *(a-c) 3D rendering of the reconstruction of the conventionally acquired tilt series of a Au@Ag nanotriangle and two central orthoslices through the reconstruction. (d-f) 3D rendering of the reconstruction of the continuously acquired tilt series and 2 central orthoslices through it. (g-i) 3D rendering of the reconstruction of the incrementally acquired tilt series and 2 central orthoslices through the reconstruction.*